\documentclass{emulateapj}

\usepackage{amsmath}
\usepackage{natbib}
\usepackage{graphics}
\usepackage{hyperref}

\newcommand{\Hb}{H{$\beta$}}
\newcommand{\Ha}{H{$\alpha$}}
\newcommand{\oiii}{[\ion{O}{3}]}
\newcommand{\msun}{\ensuremath{{\rm M}_{\odot}}}
\newcommand{\kms}{km~s\ensuremath{^{-1}}} 
\newcommand{\rth}{R\ensuremath{_{200}}}
\newcommand{\vd}{\ensuremath{\sigma_{v}}}
\newcommand{\ergs}{\text{erg~s}\ensuremath{^{-1}}}
\newcommand{\fAGN}[3]{\ensuremath{f(L_{X,#1} > 10^{#2};M_R < {#3})}}

\begin{document}
\title{THE AGN POPULATION IN X-RAY SELECTED GALAXY GROUPS AT $0.5 < z < 1.1$}

\author{Semyeong Oh\altaffilmark{1,8}, 
John S. Mulchaey\altaffilmark{2},
Jong-Hak Woo\altaffilmark{1,2,9}
Alexis Finoguenov\altaffilmark{3},
Masayuki Tanaka\altaffilmark{4},
Michael C. Cooper\altaffilmark{5},
Felicia Ziparo\altaffilmark{6},
Franz E. Bauer\altaffilmark{7},
Kenta Matsuoka\altaffilmark{1}}

\altaffiltext{1}{Astronomy Program, Department of Physics and Astronomy, Seoul National University, Seoul 151-742, Republic of Korea}

\altaffiltext{2}{Carnegie Observatories, 813 Santa Barbara Street, Pasadena, CA 91101-1292, USA}

\altaffiltext{3}{Department of Physics, University of Helsinki, Gustaf Hallstromin katu 2a, FI-00014 Helsinki, Finland}

\altaffiltext{4}{National Astronomical Observatory of Japan 2-21-1 Osawa, Mitaka, Tokyo 181-8588, JAPAN}

\altaffiltext{5}{Center for Galaxy Evolution, Department of Physics and Astronomy, University of California, Irvine, 4129 Frederick Reines Hall Irvine, CA 92697, USA}

\altaffiltext{6}{School of Physics and Astronomy, University of Birmingham, Edgbaston, Birmingham B15 2TT, UK}

\altaffiltext{7}{Pontificia Universidad Catolica de Chile, Departamento de Astronoma y Astrofsica, Casilla 306, Santiago 22, Chile}

\altaffiltext{8}{Current address: Department of Astrophysical Sciences, Princeton University, Princeton, NJ 08544, USA}

\altaffiltext{9}{Author to whom correspondence should be addressed.}

\shortauthors{OH ET AL.}

\begin{abstract}

We use Chandra data to study the incidence and properties of Active
Galactic Nuclei (AGN) in 16 intermediate redshift ($0.5 < z < 1.1$)
X-ray-selected galaxy groups in the Chandra Deep Field-South.  We
measure an AGN fraction of $\fAGN{H}{42}{-20} = 8.0_{-2.3}^{+3.0}\%$
at $\bar{z} \sim 0.74$, approximately a factor of two higher than the
AGN fraction found for rich clusters at comparable redshift. This
extends the trend found at low redshift for groups to have higher AGN
fractions than clusters.  Our estimate of the AGN fraction is also
more than a factor of 3 higher than that of low redshift
X-ray-selected groups.  Using optical spectra from various surveys, we
also constrain the properties of emission-line selected AGN in these
groups.  Contrary to the large population of X-ray AGN ($N(L_{X,H} >
10^{41}~\ergs) = 25$), we find only 4 emission-line AGN, 3 of which
are also X-ray bright.  Furthermore, most of the X-ray AGN in our
groups are optically-dull (i.e. lack strong emission-lines) similar to
those found in low redshift X-ray groups and clusters of
galaxies. This contrasts with the AGN population found in low redshift
optically-selected groups which are dominated by emission-line AGN.
The differences between the optically and X-ray-selected AGN
populations in groups are consistent with a scenario where most AGN in
the densest environments are currently in a low accretion state.
\end{abstract}

\keywords{galaxies: active --- galaxies: groups: general}

\section{INTRODUCTION}
Active Galactic Nuclei (AGN) are now known to play an integral role in
galaxy formation and evolution. Although it is established that AGN
are powered by the accretion of matter onto supermassive black holes,
the trigger mechanism(s) of such nuclear activity is still a matter of
debate \citep[e.g.,][]{choi2009, lee2012}.  The environmental
dependence of nuclear activity may provide key clues not only on
possible fueling mechanisms but also on the evolution of AGN with
changes in environment \citep[e.g.,][]{martini2007, shen2007,
  arnold2009}.

One long-standing candidate for triggering of AGN activity is major
mergers and interactions between galaxies. Suggested by earlier
studies based on the morphologies of quasar host galaxies
\citep{gehren1984, heckman1984, hutchings1992}, the merging scenario
has also been supported by numerical simulations
\citep[e.g.,][]{barnes1996}, and clustering analysis
\citep{hennawi2006, serber2006}.  On the other hand, various studies
of X-ray selected AGN highlight the importance of internal processes,
finding no enhancement of AGN activity in interacting galaxies as
predicted by merger-driven fueling (cf. Koss et al. 2011).  These
studies employ various approaches from detailed analysis of host
galaxy morphology \citep[e.g.,][]{grogin2005, garbor2009,
  cisternas2011, boehm2013, kocevski2012} to environmental analysis
\citep[e.g.,][]{silverman2009,silverman2010}.  The discrepancies
underline the complexity of the matter, including the degeneracies
induced by the correlation between galaxy and AGN properties, the need
for statistically sound control samples, the likely intrinsic
dependence on cosmic epoch, and the different physical scales of
environment probed \citep[see][for review]{martini2004}.

The incidence of AGN in galaxy groups and clusters provides yet
another observational constraint on AGN fueling mechanisms.  Naively,
there are two main prerequisites to be met for a galaxy with a
supermassive blackhole to harbor an AGN: the fuel itself and active
feeding mechanisms.  Galaxy groups are believed to be particularly
conducive to initiate nuclear activity because of the balanced
condition between the supply of cold gas and the increased chance of
gravitational interaction.  Many authors have attempted to study the
AGN populations in galaxy groups \citep{shen2007, georga2008,
  arnold2009, allevato12, tanaka12, pentericci2013} and in general
these studies are consistent with a higher AGN fraction in groups than
in clusters. This would appear to support the idea that galaxies in
groups retain larger reservoirs of cold gas to fuel AGN activity than
their counterparts in clusters.

Another important question is how AGN activity evolves in time.  There
has been considerable effort to measure the AGN fraction in groups and
clusters at higher redshifts.  \citet{eastman2007} was the first to
report a significant increase in the cluster AGN population from $z
\sim 0.2$ to $z \sim 0.6$.  \citet{martini2009} confirmed this trend
up to $z \sim 1$, while \citet{galametz2009} showed it extends up to
at least z $\sim$ 1.5.  This suggests that the AGN population in rich
clusters has experienced substantial evolution due to the cluster
environment from $z \sim 1$ to present in a manner similar to
star-forming galaxies \citep{butcher1984}.  Studies of groups suggest
a similiar trend with redshift, but with the AGN fraction in groups
being higher than clusters at all redshifts studied so far. While the
AGN fraction in the field also appears higher than in clusters at low
redshift, \citet{martini2013} find that at $1<z<1.5$ the cluster and
field AGN fractions are similar.  Observations at $z \sim 2-3$ suggest
the AGN fraction in protoclusters actually exceeds AGN fraction in the
field at these redshifts \citep{lehmer2009,digby2010}.

Here, we report the incidence and the properties of AGN in 16 X-ray
selected groups at $0.5<z<1.1$ in the Chandra Deep Field-South (CDFS).
Our sample is based on the galaxy group catalog produced by
\citet{finoguenov2014} who searched for extended X-ray emission
associated with galaxy groups.  An advantage of our study is that the
large multi-wavelength datasets in the CDFS provide multiple mass
calibrations for these groups which in turn allows us to derive
measurements of \rth{}, the radius at which the density becomes 200
times the critical density. This quantity is useful in estimating the
virialized part of groups \citep{connelly2012} and enables us to
explore possible radial trends of AGN properties. Our group sample
also benefits from the high spectroscopic completeness in the CDFS.
We combine our results with previous studies of the AGN content in
groups and clusters to investigate AGN evolution in dense
environments.  Throughout the paper, we adopt $H_0 =
70$~\kms{}~Mpc$^{-1}$, $\Omega_M = 0.3$, and $\Omega_\Lambda = 0.7$.

\begin{deluxetable*}{cccccccccc}
\tablecaption{Group Properties}

\tablehead{
\colhead{ID}  & \colhead{RA} & \colhead{DEC} &
\colhead{Redshift} & \colhead{\rth} 					& \colhead{\vd} &
\colhead{N}		& \colhead{N$_{spec}$} 					& \colhead{N$^*$}							 & \colhead{$L_{X,H}^{limit}$} \\ 
\colhead{} 		& \colhead{(deg)} 							& \colhead{(deg)} &
\colhead{} 		& \colhead{(kpc)} 							& \colhead{(\kms)} &
\colhead{} 		& \colhead{} 	& \colhead{} 	 & \colhead{($\times 10^{42}$~\ergs{})} \\
\colhead{(1)} & \colhead{(2)} & \colhead{(3)} &
\colhead{(4)} & \colhead{(5)} & \colhead{(6)} &
\colhead{(7)} & \colhead{(8)} & \colhead{(9)} & \colhead{(10)}
} 
\startdata
10 & 53.00045 & -27.90970 & 0.736498 &  459 &   446 $\pm$  38 &   9 &  9 & 7 & 0.63 \\
20 & 53.11008 & -27.67536 & 1.040575 &  500 &   518 $\pm$ 121 &   8 &  6 & 2 & 1.13 \\
21 & 53.19170 & -27.68820 & 0.731994 &  494 &   362 $\pm$  70 &  23 & 20 & 7 & 0.52 \\
24 & 52.98843 & -27.70788 & 0.666333 &  405 &   342 $\pm$ 189 &   7 &  6 & 3 & 0.56 \\
25 & 53.04006 & -27.71170 & 0.733791 &  453 &   440 $\pm$ 154 &   8 &  7 & 4 & 0.35 \\
27 & 53.21977 & -27.74213 & 0.534023 &  388 &   258 $\pm$  98 &   5 &  5 & 2 & 0.14 \\
29 & 53.07747 & -27.79237 & 0.734147 &  421 &   515 $\pm$ 148 &  16 & 16 & 7 & 0.07 \\
30 & 52.96073 & -27.82146 & 0.679769 &  559 &   492 $\pm$  77 &  19 & 15 & 7 & 0.45 \\
33 & 53.09682 & -27.82853 & 0.577203 &  337 &   259 $\pm$  58 &   4 &  4 & 1 & 0.03 \\
39 & 53.27147 & -27.86921 & 0.519614 &  385 &   494 $\pm$ 223 &   8 &  7 & 3 & 0.30 \\
41 & 53.07577 & -27.87383 & 1.096602 &  349 &   375 $\pm$ 171 &   7 &  6 & 5 & 0.40 \\
43 & 53.07750 & -27.90421 & 0.964226 &  388 &   277 $\pm$ 104 &  10 &  8 & 3 & 0.51 \\
44 & 53.02336 & -27.91458 & 0.681612 &  478 &   969 $\pm$ 276 &  11 & 11 & 4 & 0.43 \\
63 & 53.08536 & -27.74331 & 0.523297 &  378 &   327 $\pm$ 309 &   3 &  3 & 1 & 0.06 \\
68 & 53.12672 & -27.88589 & 0.645020 &  313 &   157 $\pm$  99 &   4 &  4 & 0 & 0.12 \\
76 & 53.21918 & -27.70786 & 1.029214 &  412 &  1140 $\pm$ 381 &   8 &  5 & 3 & 1.15   \enddata

\tablecomments{
	Columns: (1) Identification number; 	(2, 3) RA, DEC (J2000) of the group position defined as the center of the diffuse X-ray emission;
	(4) Group redshift;
	(5) \rth;
	(6) Velocity dispersion of the group derived using the gapper algorithm. Errors were estimated using a jackknife method.;
	(7) The number of group members with $M_R < -20$;
	(8) The number of spectroscopically confirmed group members with $M_R < -20$;
	(9) The number of group members with $M_R < M_R^* + 1$;
	(10) The X-ray point source luminosity detection limit
}

\label{tab:groups}
\end{deluxetable*}

\section{DATA \& SAMPLE}

Our groups are taken from the catalog given in \citet{finoguenov2014}.
To identify groups, deep images of the CDFS with Chandra
\citep{xue2011} and XMM-Newton \citep{comastri2011} were used to
search for extended X-ray emission from the intragroup medium.  To
detect extended emission, we first remove the flux from point sources
following the proceedure outlined in \citet{fin09,fin10}. Point
sources are removed separately from the XMM-Newton and Chandra
datasets to allow for AGN variability and astrometry differences. The
two datasets are combined after the point sources and background has
been removed. For the identification of possible galaxy groups, we
restrict our analysis to X-ray sources that are extended on a spatial
scale of 32$''$ or greater.

To verify the emission is associated with a group, we require the
system to be detected in optical/near-infrared images.  A red sequence
finder is used to help identify optical counterparts.  We further
require that the system be spectroscopically confirmed using the
existing redshifts from the CDFS. This analysis results in a sample of
46 X-ray groups spanning a large range of X–-ray luminosities
(10$^{41}$ -− 10$^{43}$ ergs s$^{−1}$).  A small number of these
groups have already been studied by other authors \citep{tanaka2013,
  ziparo2013}.  Here, we focus on 16 groups between 0.5 $< z <$ 1.1
that are both detected with high confidence as extended sources in the
Chandra images and have good membership data available.  We note that
three (29, 41, 43) of the groups studied here are in common with the
sample of \citet{pentericci2013} (GS~4, GS~8, GS~5). However, only
GS~5 was identified as an X-ray group in their work.

With the location and the approximate redshift of the groups in hand,
group members were identified by the method outlined in
\citet{mulchaey2006}.  For this, we use a compilation of spectroscopic
redshifts compiled by the ESO-GOODS team ({\tt MASTERCAT\_v2.0}) with
additions from \citet{silverman2010} and the Arizona CDFS Environment
Survey \citep{cooper2012}.  We start by considering all galaxies
within $\pm 2000$~\kms{} from the approximate redshift of the group,
and within a radial distance of \rth{} from the center of the
group. From these galaxies, we determine the redshift ($\bar{z}$) and
velocity dispersion ($\vd$) of the group.  If a galaxy has a radial
velocity that is more than $3\vd$ from the new redshift of the group,
it is excluded from the group and a new mean redshift and velocity
dispersion is calculated.  We iterate this process until no further
change in group membership is made.  In each step, the velocity
dispersion ($\sigma_v$) of each group is determined using the gapper
algorithm \citep{beers1990}. This process resulted in the elimination
of no more than 1 potential group member from any of the groups in our
sample.  The final group properties are listed in
Table~\ref{tab:groups}.  The dynamical mass of groups in our sample
ranges from $\log{M/\msun} = 12.7$ to $\log{M/\msun} = 14.5$ with the
median of $\log{M/\msun} = 13.7$.

We assess the spectroscopic completeness of our group membership using
the MUSYC photometric redshift catalog \citep{cardamone2010}.  We
apply the same criteria above, and correct the size of each group with
the number of additional potential group members found.  For each
group, we list this corrected size (N), and the size only including
galaxies with spectroscopic redshift (N$_{spec}$) in column 7 and 8 of
Table~\ref{tab:xraysrc}.  In total, we find 18 additional galaxies
that are potential members of these groups, achieving a very high
(88\%) spectroscopic completeness for galaxies more luminous than
$M_R<-20$.  Given this high completeness, any missed group members
will have a negligible impact on our calculation of the AGN fraction
since the Poisson error on the number of AGN dominates the error.

To find X-ray AGN, we match group members with the 4Ms CDFS Source
Catalog \citep{xue2011}.  For direct comparison with
\citet{eastman2007}, we identify X-ray point sources with rest-frame
hard (2-10 keV) X-ray luminosity $L_{X,H}>10^{42}$~\ergs{} as AGN.  We
assume a power-law photon index of 1.8 to convert Chandra
absorption-corrected full band luminosity to $L_{X,H}$.  Twenty five
X-ray point sources are found. All but one of these X-ray sources is
associated with a galaxy with a confirmed spectroscopic redshift.  The
X-ray sources matched with group members and their properties are
listed in Table~\ref{tab:xraysrc}.

We also identify emission-line AGN using archival spectra from the
literature.  Typically, separating AGN from star-forming galaxies
requires measurements of the emission line flux ratio \oiii{}/\Hb{}
and [\ion{N}{2}]/\Ha{}.  In our case, this is not applicable since the
wavelength range including \Ha{}--[\ion{N}{2}] is not accessible with
the existing spectra for our groups.  Instead, we use the
classification scheme by \citet{yan2011}, which mitigates the problem
by using the rest-frame $U-B$ color in place of the [\ion{N}{2}]/\Ha{}
ratio.  For all group members with a spectroscopic redshift, we look
for archival optical spectra from which the redshift was measured.  We
found spectra for 104 sources (70\%), 68 of which cover the
H$\beta$--\oiii{} wavelength range.  When both lines are detected, we
measure the \oiii{}/\Hb{} line ratio by fitting a Gaussian to each
line after a polynomial continuum level is subtracted.  Otherwise, we
infer a $3\sigma$ limit from the noise level adjacent to the line of
interest.  The rest-frame $U-B$ colors are measured using EAZY
\citep{brammer2008} with redshifts fixed to those determined from
spectroscopy.

Finally, we use the existing optical spectra to study the
emission-line properties of our X-ray selected AGN.  Only two out of
twelve X-ray AGN with archival spectra covering \Hb-\oiii{} show
emission lines. Therefore, we find that most of the X-ray AGN in our
group sample are \lq\lq optically-dull\rq\rq\, similar to those found
in rich clusters \citep{eastman2007}.

\begin{deluxetable}{ccccccc}
	\tablecaption{X-ray Sources in Groups}
	\tablehead{
		\colhead{XID} & \colhead{RA} & \colhead{DEC} & 
		\colhead{Z} & \colhead{Group} & \colhead{$\log{L_{X,H}}$}\\ 
		\colhead{(1)} & \colhead{(2)} & \colhead{(3)} & \colhead{(4)} &
		\colhead{(5)} & \colhead{(6)} 
	}
	\startdata	
	355 & 53.10779 & -27.68006 & 1.045020 &     20 &       42.27 \\ 
	367 & 53.11046 & -27.67664 & 1.043430 &     20 &       43.99 \\ 
	601 & 53.18929 & -27.68250 & 0.734000\tablenotemark{a} &     21 &       42.04 \\ 
	627 & 53.19967 & -27.69675 & 0.735700 &     21 &       43.49 \\ 
	581 & 53.18300 & -27.70061 & 0.731780 &     21 &       41.73 \\ 
	 63 & 52.99063 & -27.70242 & 0.667960 &     24 &       42.64 \\ 
	 62 & 52.98987 & -27.71442 & 0.666000 &     24 &       42.11 \\ 
	146 & 53.03825 & -27.70722 & 0.734000 &     25 &       42.39 \\ 
	148 & 53.03929 & -27.70994 & 0.729000 &     25 &       41.93 \\ 
	151 & 53.04054 & -27.71347 & 0.735040 &     25 &       41.98 \\ 
	198 & 53.05775 & -27.71369 & 0.734550 &     25 &       43.14 \\ 
	654 & 53.21687 & -27.74014 & 0.534210 &     27 &       41.55 \\ 
	647 & 53.21379 & -27.74158 & 0.533360 &     27 &       41.58 \\ 
	250 & 53.07517 & -27.78856 & 0.734320 &     29 &       41.40 \\ 
	209 & 53.06183 & -27.79408 & 0.735000 &     29 &       41.34 \\ 
	246 & 53.07346 & -27.80336 & 0.734000 &     29 &       41.66 \\ 
	352 & 53.10729 & -27.82686 & 0.575980 &     33 &       41.63 \\ 
	322 & 53.09863 & -27.82736 & 0.578540 &     33 &       40.83 \\ 
	712 & 53.27346 & -27.87056 & 0.521200 &     39 &       42.04 \\ 
	243 & 53.07163 & -27.87250 & 1.095610 &     41 &       44.04 \\ 
	269 & 53.08029 & -27.90189 & 0.964000 &     43 &       42.09 \\ 
	116 & 53.02492 & -27.91397 & 0.684600 &     44 &       41.93 \\ 
	292 & 53.08871 & -27.74336 & 0.523480 &     63 &       41.18 \\ 
	423 & 53.12567 & -27.88500 & 0.645050 &     68 &       41.45 \\ 
	652 & 53.21608 & -27.70825 & 1.022730 &     76 &       42.72    
	\enddata
	\tablecomments{Columns:
		(1) Chandra 4Ms Source ID;
		(2, 3) RA, DEC;
		(4) Redshift;
		(5) ID of the matched group (see Table~\ref{tab:groups});
		(6) Hard band X-ray luminosity;
	}
	\tablenotetext{a}{photometric redshift}
	\label{tab:xraysrc}
\end{deluxetable}

\begin{figure*}[htb]
  \begin{center}
    \includegraphics[width=0.7\textwidth]{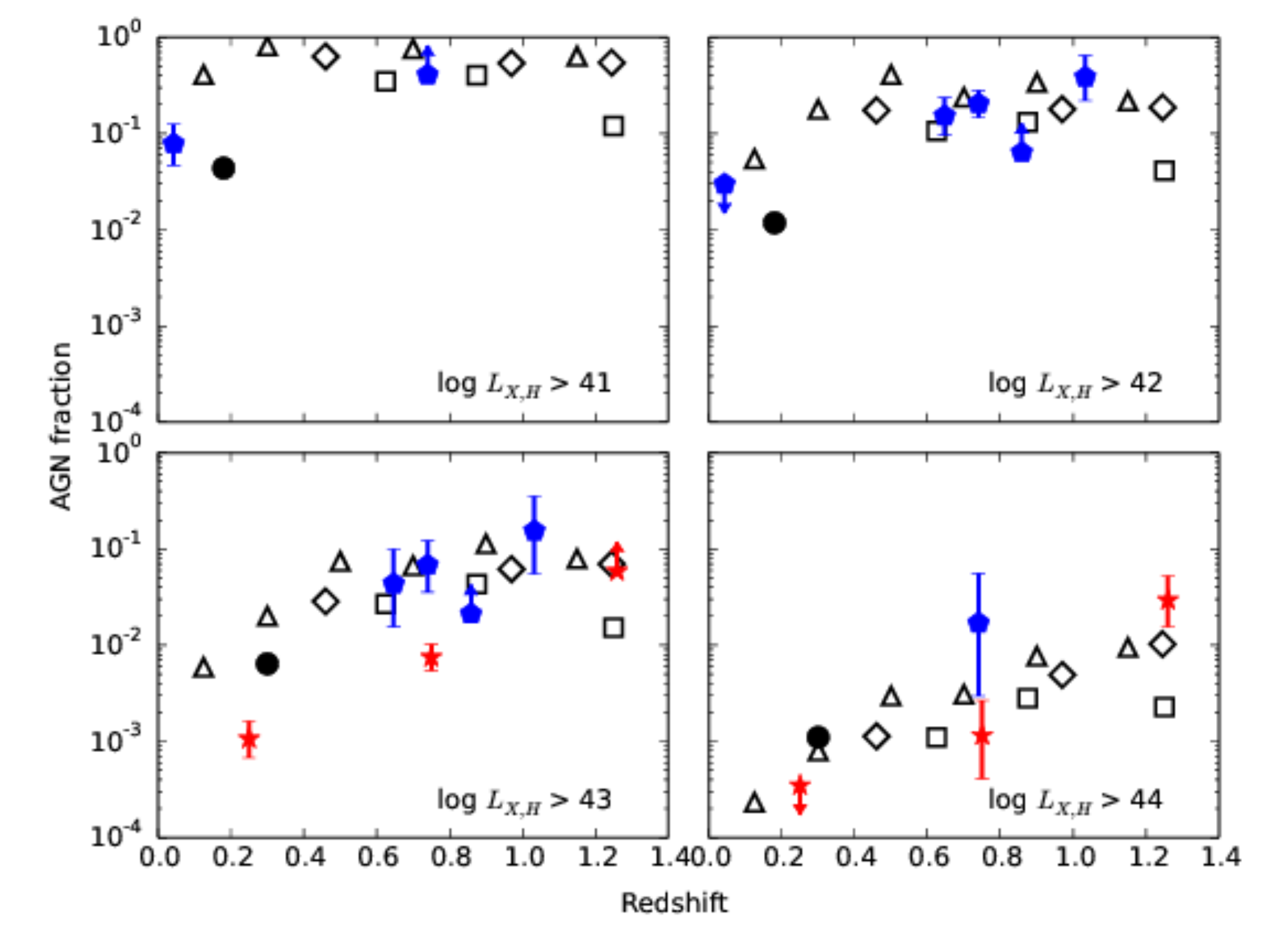}
  \end{center}
  \caption{Evolution of the AGN fraction in groups (blue pentagon),
    clusters (red star), and field (black circle).
    The different panels correspond to different X-ray luminosity thresholds.
    We divide our sample into
    low ($\bar{z} = 0.647$) and high ($\bar{z} = 1.033$) redshift bins.
    AGN fractions for low redshift groups are taken from \citet{arnold2009}, 
    \citet{pentericci2013} for groups at $z \sim 0.8$, and
    \citet{martini2013} for all clusters.
    We also plot the field AGN fraction from \citet{haggard2010} at $z \sim 0.2$
    (from their Table~4, $f(L_{X,0.5-8~keV} > 10^{41}/10^{42};M_R < -20)$),
    and at $z \sim 0.3$ as reported by \citet{martini2013}.
    Field AGN fractions calculated using the field hard X-ray luminosity function
    \citep{ueda2003} and the galaxy luminosity function from
    \citet{ilbert2005} (empty circle), \citet{dahlen2005} (empty triangle), and
    \citet{chen2003} (empty square) are also included.}
  \label{fig:zevol}
\end{figure*}

\section{RESULTS}
\subsection{The Cosmic Evolution of the AGN Fraction\\in Groups and Clusters}

The determination of the AGN fraction critically depends on the AGN
classification scheme and the host galaxy luminosity threshold
adopted. Here, we make direct comparisons with previous studies by
adopting the same AGN and galaxy luminosity thresholds used in the
earlier studies.  Through out the paper, the errors quoted for the AGN
fractions correspond to the $1\sigma$ (68\%) confidence interval of
the Poisson distribution.

First, we compare our result with \citet{eastman2007}, who reported
$\fAGN{H}{42}{-20} = 2.8_{-1.0}^{1.5}\%$ from a sample of four galaxy
clusters at $z \sim 0.6$.  We find 25 X-ray sources out of 150 group
members, 12 of which have $L_{X,H}>10^{42}$~\ergs{}, yielding an AGN
fraction $\fAGN{H}{42}{-20} = 8.0_{-2.3}^{+3.0}\%$.  Thus, our
estimate for groups at $\bar{z} \sim 0.74$ is larger than that for
clusters by more than a factor of 2. This factor is consistent with
the one found for groups and clusters at $z \sim 0.043$
\citep{arnold2009}.  If we only consider the 12 groups at $z<0.8$
$(\bar{z}=0.65)$ in our sample to better match the redshift of the
cluster sample, we find the trend is weakened but still present.

We also directly compare our result with low redshift X-ray groups
\citep{arnold2009}.  To match the low redshift sample, we limit our
sample to groups with $\sigma < 500$~\kms{}.  After converting
$L_{X,H}$ to $L_{0.3-8~keV}$ using a power-law photon index of 1.8, we
find $\fAGN{0.3-8~keV}{41}{-20} = 16.8_{-3.9}^{+4.9}\%$ (18/107), a
factor of 3 larger than the $5.5_{-1.8}^{+2.5}\%$ (9/164) found for
low redshift X-ray groups \citep{arnold2009}.  Note that our estimate
is a lower limit to the true value since in many of our groups, we do
not reach the depth of $L_{X,H}=10^{41}\ergs$ used in
\citet{arnold2009} (Table~\ref{tab:xraysrc}).  Therefore, like in rich
clusters \citep{martini2013}, the AGN fraction in X-ray groups drops
significantly from z $\sim$ 1 to the present day.

We also note that our estimate for the AGN fraction in groups is
consistent with \citet{pentericci2013} which recently reported
$\fAGN{H}{42}{-20} = 6.3_{-1.6}^{+2.1}\%$ (15/237) using a sample of
11 groups in the GOODS fields at $\bar{z} = 0.859$. Five of the groups
in the \citet{pentericci2013} sample are selected from the CDFS. As
noted earlier, three of these groups are in fact X-ray detected and in
our sample. Although the \citet{pentericci2013} sample was not X-ray
selected, the high fraction of overlap between our samples in the CDFS
field suggests that their detection method is also largely identifying
more massive and evolved groups.

In Figure~\ref{fig:zevol}, we show the evolution of the AGN fraction
in groups and clusters.  To properly compare the AGN fraction across
redshifts, we need to account for the evolution of the galaxy
population.  Following \citet{martini2009}, we adopt an evolving
galaxy luminosity threshold $M_R < M_R^*(z) + 1$ where $M_R^*(z) =
-21.92 - z$.  For clusters, we use values from \citet{martini2013}
binned at $z = 0.25, 0.75$, and $1.26$ which includes data from
\citet{eastman2007}.  As can be seen from Figure~\ref{fig:zevol}, the
frequency of AGN activity is enhanced in groups over clusters by
approximately an order of magnitude for $L_{X,H} > 10^{43} \ergs$.  In
addition, we find that the AGN fraction in groups increases with
redshift.  This result is consistent with the trend found for clusters
and verifies the result reported by \citet{pentericci2013} for groups.

\citet{haggard2010} investigated the AGN fraction in the field to
$z=0.7$ using the {\it Chandra} Multiwavelength Project and the Sloan
Digital Sky Survey.  By matching AGN X-ray luminosity and host galaxy
R-band magnitude threshold with those of \citet{martini2007}, they
find an excellent agreement between the field and cluster AGN fraction
at low redshift ($0.05 < z < 0.31$) for relatively low luminosity AGNs
($L_{X,H} \sim 10^{41-42}\ergs$).  A somewhat different picture
emerges for higher luminosity AGNs ($L_{X,H} \sim 10^{43-44}\ergs$).
For these objects, \citet{martini2013} found that the field AGN
fraction is about six times higher than in clusters at low
redshift. However, the fractions are comparable in these environment
even for luminous AGN by $z\sim 1.25$ \citep{martini2013}.  Here, we
find that at $z\sim 0.7$ the AGN fraction for higher luminosity AGNs
in groups is significantly higher that that in clusters.  Currently,
the field AGN fraction based on the selection criteria adopted in
  this paper at this redshift is not well-constrained and it is
unclear whether the AGN fraction in groups is higher than that of
field.  As a first attempt, we follow \citet{martini2009}, and
  calculate the ``field'' AGN fraction using the field hard X-ray
  luminosity function of \citet{ueda2003} and the galaxy luminosity
  functions from the VIMOS-VLT Deep Survey \citep{ilbert2005}, the Las
  Campanas Infrared Survey \citep{chen2003}, and the Great
  Observatories Origins Deep Survey \citep{dahlen2005}.  We normalize
  the number density of hard X-ray sources given from the
  luminosity-dependent density evolution model by the number density
  of galaxies brighter than $M_R^*(z) +1$ where $M_R^*(z)$ is given by
  each luminosity function.  We find that the AGN fraction of our
  group sample is in general consistent with the field AGN fractions
  from luminosity functions at all redshifts and luminosity
  thresholds.  However, we caution that these field estimates may
  be contaminated by galaxies in groups. Nonetheless,
the general picture emerges that high luminosity AGNs ($L_{X,H} \sim
10^{43-44}\ergs$) are sensitive to the environment and that the group
environment is more favorable for galaxies to host high luminosity
AGNs than richer environments like clusters.

\subsection{AGN Properties}

In this section, we investigate various properties of AGN in our
sample, and compare them with previous studies.  First, we compare the
X-ray and emission-line AGN content of groups and clusters.  In
contrast to the $> 60$\% of galaxies in low redshift poor groups
showing strong emission lines \citep{shen2007}, only 8 of 68 galaxies
(18\%) in our groups with optical spectra covering \Hb-\oiii\, show
measurable emission lines.  Of these 8 galaxies, only one galaxy has
line ratios consistent with hosting a narrow-line AGN according to the
\citet{yan2011} diagnostics, and this object also has a high X-ray
luminosity ($L_{X,H} = 1.4\times10^{43}$~\ergs{}).  In addition, there
are three more AGN candidates detected by their broad \ion{Mg}{2}
emission in the near-UV or their broad band SEDs.  In total, four
strong emission-line AGN are found in our group sample, three of which
are also X-ray bright.  As noted earlier, most of the X-ray AGN in our
sample are optically dull similar to the nearly disjoint population of
optical and X-ray AGN in X-ray groups at low redshift
\citep{arnold2009} or X-ray AGN in clusters at similar redshift
\citep{eastman2007}.  On the other hand, emission-line AGN appear to
be more common than X-ray bright, optically-dull AGN in groups that do
not contain a significant intragroup medium component
\citep{shen2007}.  These results are consistent with the idea that
X-ray bright, optically-dull AGN are more common in more virialized
systems, as suggested by some accretion evolution scenarios
\citep{shen2007}.

\begin{figure}[thbp]
  \centering
    \includegraphics[width=0.50\textwidth]{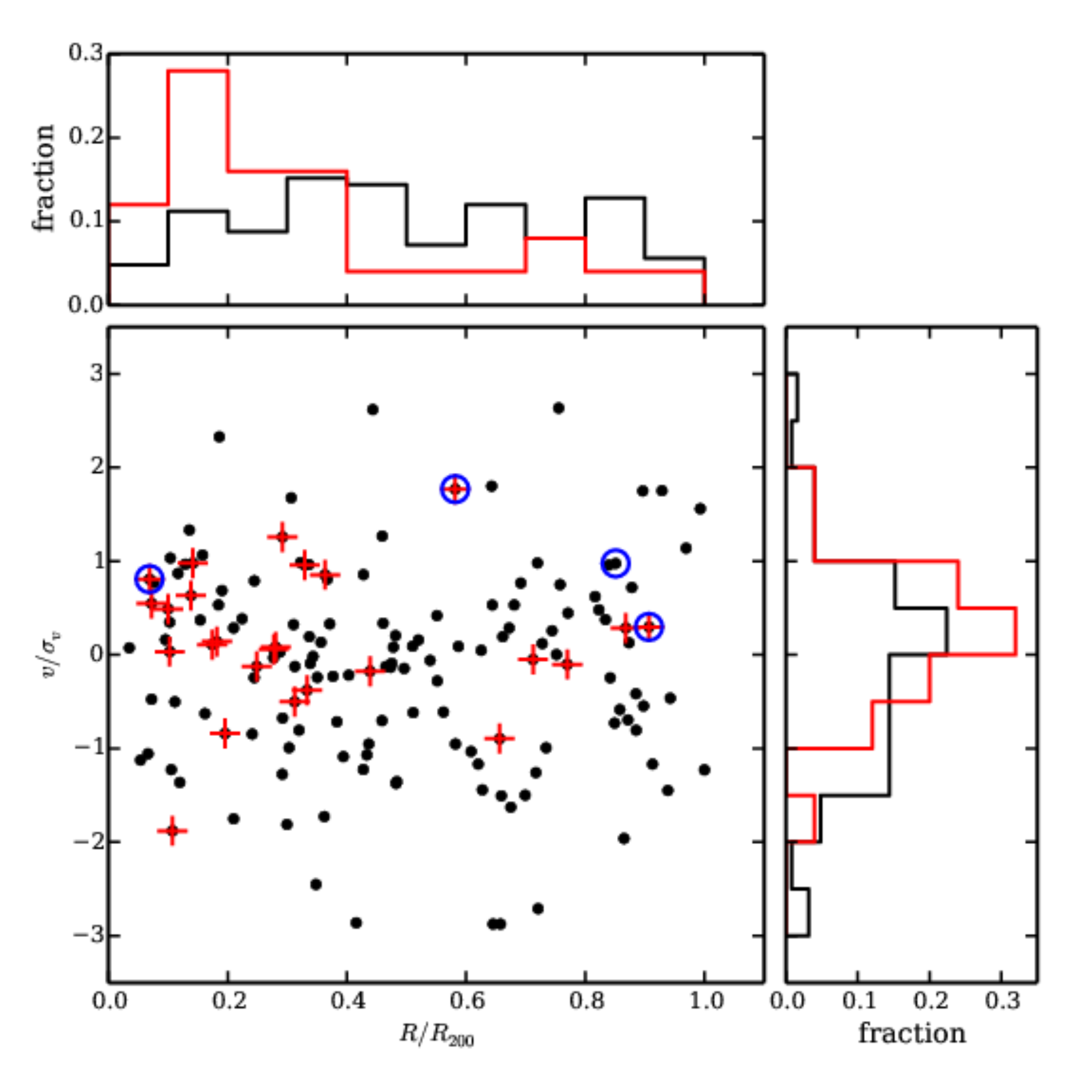}          
    \caption{Distribution of X-ray AGN (red crosses) and non-AGN galaxies (black points)
    in the $r/\rth - v/\vd$ plane (lower left).
    Shared axis plots show the marginalized distribution of each parameter
    for X-ray AGN (red) and non-AGN galaxies (black).
    Optical AGN are enclosed in blue circles.
    }
  \label{fig:dist}
\end{figure}

\begin{figure*}
  \centering
  \includegraphics[width=0.85\textwidth]{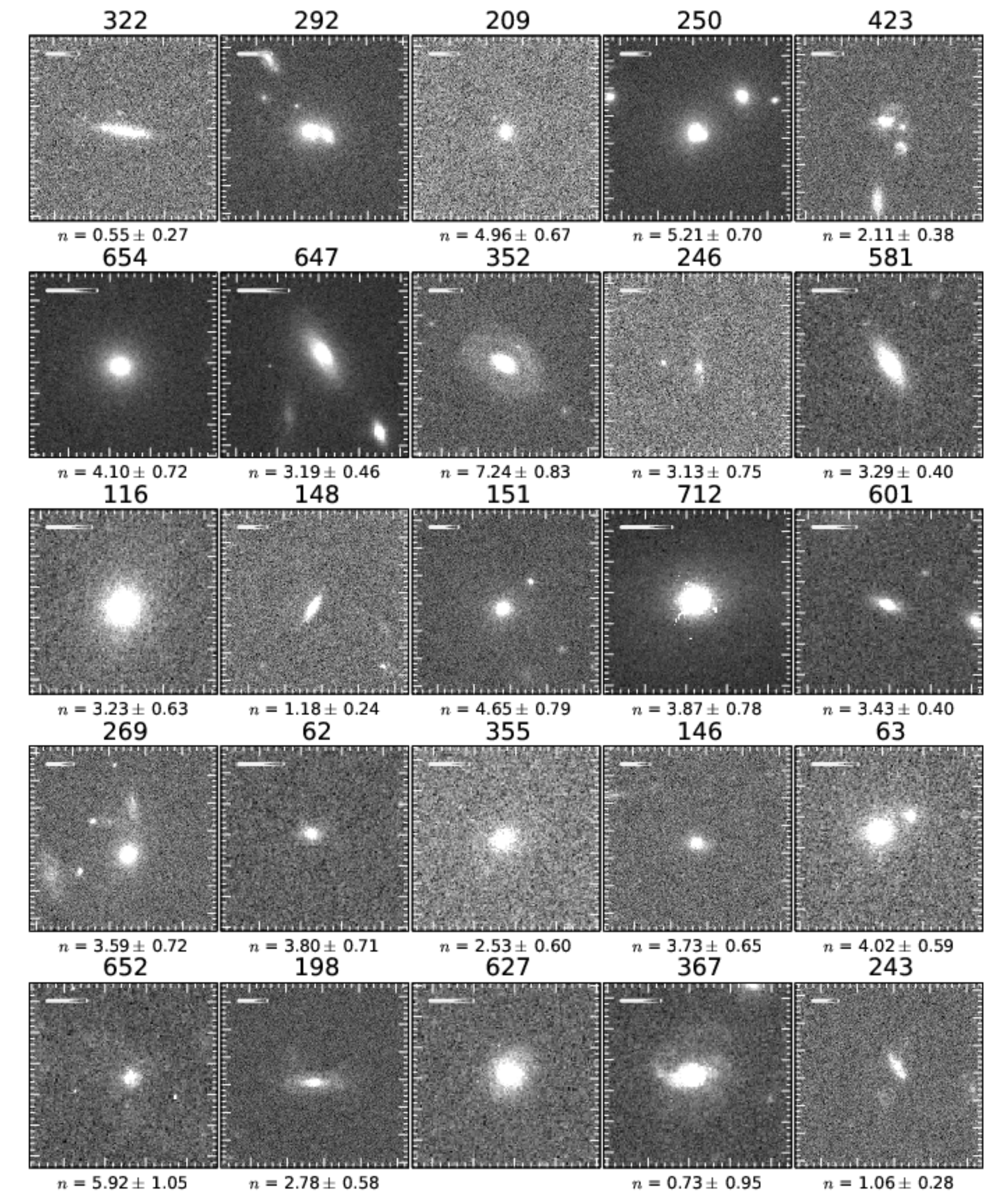}
  \caption{GEMS HST z-band images of 25 X-ray AGN ($L_{X,H}>10^{42}$~\ergs{})
    from our group sample. Each image covers $6\times6$ arcsec$^{2}$.
    Scale bars correspond to a physical scale of 10~kpc in each image. 
    The Sersic index from \citet{haussler2007} is presented at the bottom of 
    each panel. 
     }
  \label{fig:zband}
\end{figure*}

We are also interested in determining if X-ray AGN have a different
spatial distribution in groups than non-AGN hosting galaxies.  To
determine this, in Figure~\ref{fig:dist} we show the distribution of
non-AGN galaxies and galaxies hosting an X-ray source in a
groupcentric distance ($r/\rth$)--velocity ($v/\vd$) diagram.  The
velocities are normalized by the rest-frame velocity dispersion of
each group.  The marginalized distribution of each parameter is also
presented in a separate panel respectively for X-ray AGN and non-AGN
galaxies.  As can be seen from Figure~\ref{fig:dist}, X-ray sources
are more centrally concentrated than non-AGN galaxies: 75\% of all
X-ray sources, and 80\% of all X-ray AGN ($L_{X,H} > 10^{42}$~\ergs)
reside within a half of the \rth.  Our result is consistent with the
central concentration of X-ray sources in clusters
\citep{ruderman2005,martini2007,galametz2009,bignamini2008}.  However,
our result appears inconsistent with the result of
\citet{pentericci2013} who found that X-ray AGN are preferentially
located at the edge of groups and small clusters at similar redshifts.
The difference may be due to the different methods used to define the
center and extent of each group.  In the three groups in common with
their sample, we find that the center determined as the peak position
of the density distribution of each structure by their algorithm can
be offset by as much as $\sim 23$\arcsec{} from our estimate
determined from the X-ray emission.  Such a difference may have washed
out the trend we see in Figure~\ref{fig:dist} in the
\citet{pentericci2013} study.  We find no difference in the velocity
structure of AGN and non-AGN hosting galaxies in groups, consistent
with \citet{pentericci2013}.  We also find no correlation between
X-ray luminosity and groupcentric radial distance or velocity.

\subsection{Host Galaxy Properties}

The morphologies of AGN host galaxies can provide valuable information
on possible fueling mechanisms of nuclear activity.  While such
studies are often prohibited or biased by the presence of a bright
nucleus in QSO host galaxies, most of the X-ray AGN in our sample are
optically dull, making it easier to study the morphology of the
underlying stellar population.  In Figure~\ref{fig:zband} we present
the HST z-band images of all 25 X-ray sources in our group sample
using the Galaxy Evolution from Morphologies and SEDs survey
\citep[GEMS;][]{rix2004}.  Visual inspection of the X-ray detected
galaxies reveal a variety of morphologies.  While several galaxies
show evidence of nearby companions or merging signatures, the majority
shows no clear sign of major merger. Thus, it seems unlikely that the
current AGN activity in these galaxies has been triggered by an
on-going major merger.  This is consistent with the recent findings
that most of moderate-luminosity AGN are not hosted by starbursting
major mergers (e.g., Xue et al. 2010; Mullaney et al. 2012).

In Figure~\ref{fig:zband} we also list the Sersic index on the bottom
of each panel if available. These indices are derived from fitting a
single Sersic profile component to the HST z-band images
\citep[see][]{haussler2007}. Most of the host galaxies have high
Sersic indices with mean $n=3.40\pm1.59$. Although a single Sersic fit
is a somewhat limited way to characterize the complex nature of host
galaxies, it does suggest that these galaxies are mostly
bulge-dominated.  While only one (XID 292) out of 25 X-ray AGN host
galaxies is clearly a major merger, this together with the high Sersic
indices may indicate that minor mergers or internal processes may have
triggered the nuclear activity in these galaxies as minor mergers can
also make a significant contribution to the formation of bulges
\citep{hopkins2010a}.

To investigate whether the AGN activity in groups is correlated with
star formation, we use 100 and 160 $\mu$m flux measurements from the
GOODS-Herschel Survey \citep{elbaz2011} and PACS Evolutionary
Programme \citep{lutz2011}. We find nine and seven detections,
respectively at 100 and 160 $\mu$m among our 25 X-ray AGN, and compare
them with the X-ray luminosities in Figure 4.  We find that compared
to low-redshift optical or X-ray selected AGN (e.g., Netzer et
al. 2007), the far-infrared (FIR) luminosity is higher at a given AGN
luminosity, indicating that AGN in galaxy groups at $<z>=0.7$ are
hosted by galaxies with higher star formation rates than local
galaxies. On the other hand, we find no strong correlation between
X-ray and FIR luminosity although the FIR-detected sample is
small. These results are consistent with the findings of other studies
of X-ray detected galaxies.  For example, \citet{mullaney2012a} report
that moderate-luminosity X-ray AGN (log L$_X$= 42-44 \ergs) at similar
redshifts show no correlation between star formation and AGN
luminosity and that the star formation rate of X-ray AGN host galaxies
increases strongly from z$\sim$0 to 3 (see their Figure 5).  This may
indicate that X-ray AGN host galaxies in the group environment follow
a similar trend of a relative growth of stars and supermassive black
holes to the general galaxy population in other environments at the
same redshift while the higher X-ray AGN fraction in group galaxies
implies that the triggering of AGN and star formation is more frequent
in the group environment than in the cluster environment. The higher
AGN fraction is presumably due to the larger gas content and favorable
conditions for dynamical interaction among member galaxies.

\begin{figure}
  \centering
  \includegraphics[width=0.45\textwidth]{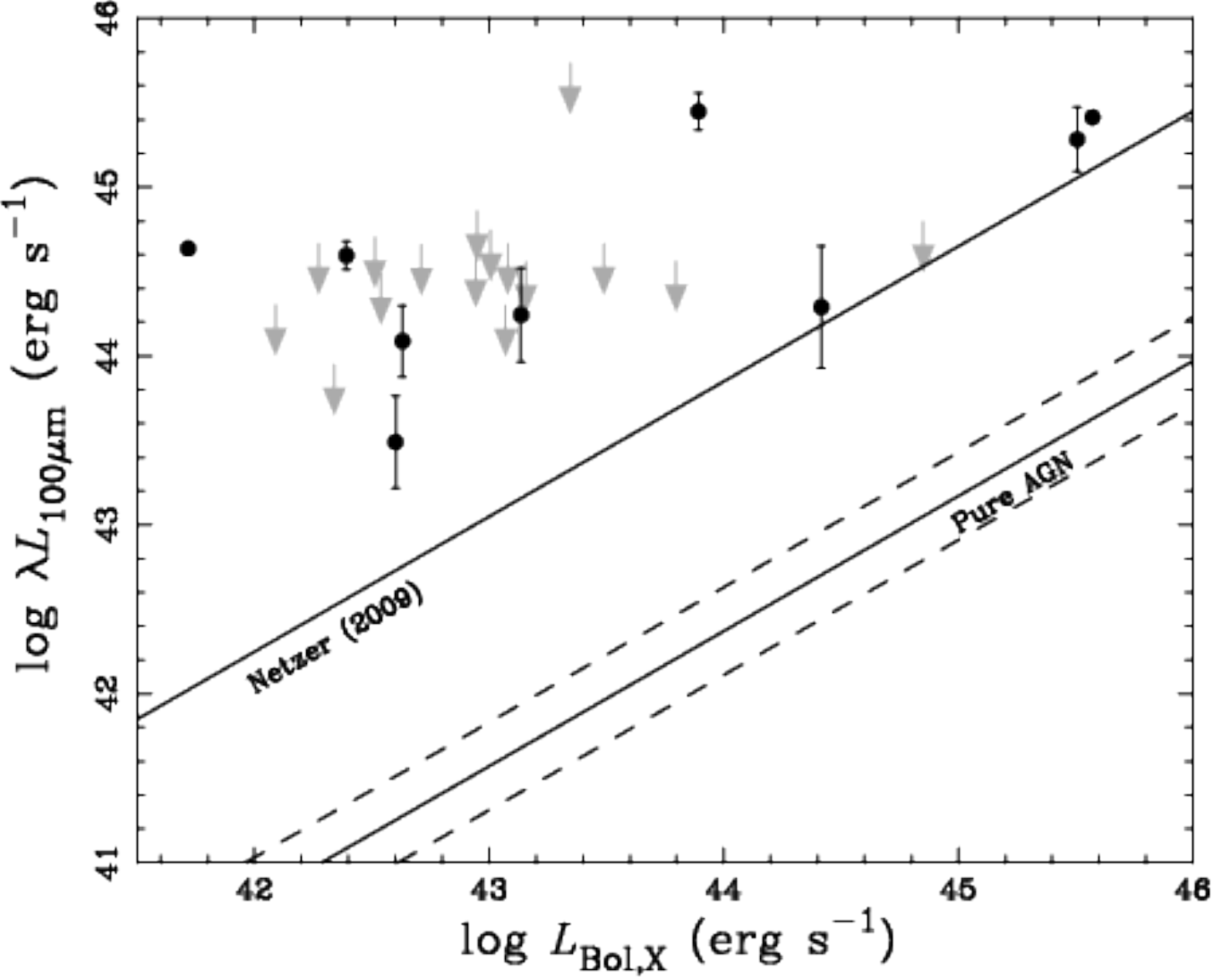}\\
  \includegraphics[width=0.45\textwidth]{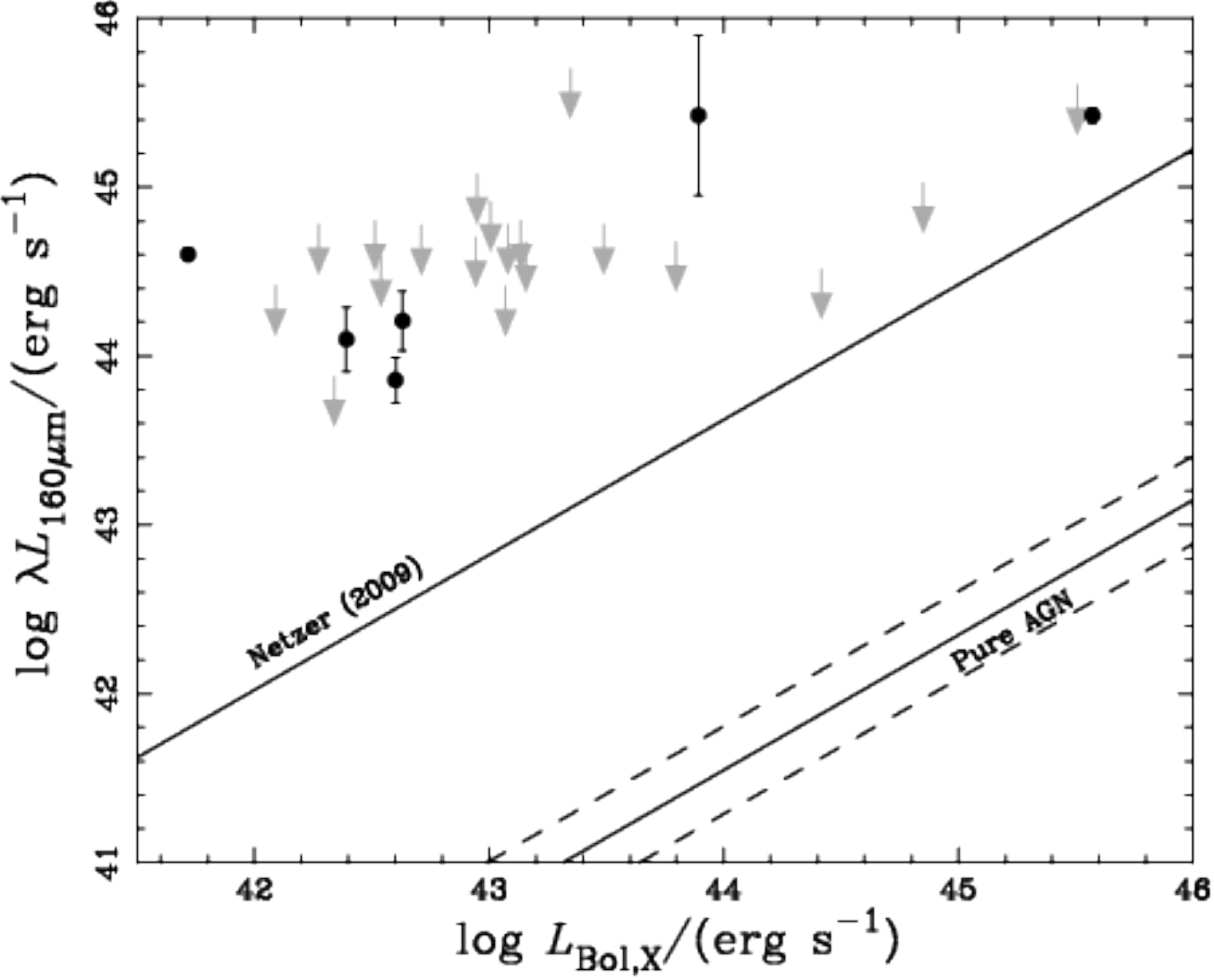}
  \caption{Comparison of X-ray based AGN luminsosity with 100 $\mu$m
    (top) and 160 $\mu$m luminosities (bottom).  Although the far-IR
    detections are limited to a small fraction of the sample (filled
    circles), no strong correlation is present between X-ray and FIR
    luminosity. The local relation (solid line) is taken from Netzer
    et al. (2009) after converting the star formation rate into far-IR
    luminosity. Note that AGN luminosity has been derived from X-ray
    luminosity using the bolometric correction method (Rigby et
    al. 2009).  The locus of AGN host galaxies without star formation
    (pure AGN) is also indicated by a solid line with 1-$\sigma$
    dispersion (Mullaney et al. 2011; Rosario et al. 2012).  }
\end{figure}

\section{SUMMARY}

We have studied the X-ray and optical AGN population in 16
X-ray-selected groups at $0.5 < z < 1.1$ in the Chandra Deep Field
South.  Our sample benefits from the extensive analysis of deep X-ray
images of CDFS, and also from the rich archival data available from
various sources. Specifically, we achieve high spectroscopic
completeness which rules out false positive assignments of AGN to
groups.  We summarize our results as follows:

\smallskip

\begin{itemize}
  {\item We measure an X-ray AGN fraction of $\fAGN{H}{42}{-20} =
    8.0_{-2.3}^{+3.0}\%$ at $\bar{z} \sim 0.74$. This AGN fraction is
    higher than that found for similar groups at low redshift.}
  {\item Similar to the results found at low redshift, the AGN
    fraction is higher in groups than in clusters at intermediate
    redshifts. This result is consistent with the expectation that
    galaxies in groups contain more cold gas and are more likely to
    experience an interaction than galaxies in rich clusters.}
		
	{\item Most of the X-ray detected AGN in these groups are
          optically dull (i.e. do not contain strong emission lines)
          similar to those found in rich clusters.}

	{\item We find that X-ray AGN are more centrally concentrated
          in groups than non-AGN hosting galaxies.  However, we find
          no apparent trend between the X-ray luminosity of the AGN
          and the group-centric radial distance or velocity.}

{\item The majority of the X-ray AGN host galaxies are bulge-dominated
  systems. Only one of the 25 galaxies shows clear evidence of an
  on-going major merger.}

   {\item By comparing the far-infrared and X-ray luminosities of
     X-ray detected galaxies, we find that star formation rate for
     given AGN luminosity is higher than in local AGN host galaxies.
     This trend is similar to that of X-ray AGN host galaxies in
     general (Mullaney et al. 2012), suggesting that the relative
     growth of stars and supermassive black holes in the group
     environment is similar to that in other environments.  }
\end{itemize}

\acknowledgements We thank the anonymous referee for a very helpful
report that helped improve this paper.  This research has been
supported by TJ Park Science Fellowship of POSCO TJ Park Foundation.
J.S.M acknowledges partial support for this work from SAO grant
SP1-12006A and HST grant AR-12831 J.H.W acknowledges the support by
the National Research Foundation of Korea (NRF) grant funded by the
Korea government (No. 2012-006087). AF acknowledges support from SAO
grant SP1-12006B.  M.T acknowledges support by KAKENHI No. 23740144.
This research made use of Astropy (\url{http://www.astropy.org}), a
community-developed core Python package for Astronomy and TOPCAT
(\url{http://www.starlink.ac.uk/topcat/}).


\begin{thebibliography}{48}
\expandafter\ifx\csname natexlab\endcsname\relax\def\natexlab#1{#1}\fi

\bibitem[Allevato et al.(2012)]{allevato12} Allevato, V., 
Finoguenov, A., Hasinger, G., et al.\ 2012, \apj, 758, 47

\bibitem[{{Arnold} {et~al.}(2009){Arnold}, {Martini}, {Mulchaey}, {Berti}, \&
  {Jeltema}}]{arnold2009}
{Arnold}, T.~J., {Martini}, P., {Mulchaey}, J.~S., {Berti}, A., \& {Jeltema},
  T.~E. 2009, \apj, 707, 1691

\bibitem[{{Barnes} \& {Hernquist}(1996)}]{barnes1996}
{Barnes}, J.~E., \& {Hernquist}, L. 1996, \apj, 471, 115

\bibitem[{{Beers} {et~al.}(1990){Beers}, {Flynn}, \& {Gebhardt}}]{beers1990}
{Beers}, T.~C., {Flynn}, K., \& {Gebhardt}, K. 1990, \aj, 100, 32

\bibitem[{{Bignamini} {et~al.}(2008){Bignamini}, {Tozzi}, {Borgani}, {Ettori},
  \& {Rosati}}]{bignamini2008}
{Bignamini}, A., {Tozzi}, P., {Borgani}, S., {Ettori}, S., \& {Rosati}, P.
  2008, \aap, 489, 967


\bibitem[B{\"o}hm et al.(2013)]{boehm2013} B{\"o}hm, A., Wisotzki, L., Bell, E.~F., et al.\ 2013, \aap, 549, A46 

\bibitem[{{Brammer} {et~al.}(2008){Brammer}, {van Dokkum}, \&
  {Coppi}}]{brammer2008}
{Brammer}, G.~B., {van Dokkum}, P.~G., \& {Coppi}, P. 2008, \apj, 686, 1503

\bibitem[{{Butcher} \& {Oemler}(1984)}]{butcher1984}
{Butcher}, H., \& {Oemler}, Jr., A. 1984, \apj, 285, 426

\bibitem[{{Cardamone} {et~al.}(2010){Cardamone}, {van Dokkum}, {Urry},
  {Taniguchi}, {Gawiser}, {Brammer}, {Taylor}, {Damen}, {Treister}, {Cobb},
  {Bond}, {Schawinski}, {Lira}, {Murayama}, {Saito}, \&
  {Sumikawa}}]{cardamone2010}
{Cardamone}, C.~N., {van Dokkum}, P.~G., {Urry}, C.~M., {et~al.} 2010, \apjs,
  189, 270

\bibitem[Chen et al.(2003)]{chen2003} Chen, H.-W., Marzke, 
R.~O., McCarthy, P.~J., et al.\ 2003, \apj, 586, 745 

\bibitem[{{Choi} {et~al.}(2009){Choi}, {Woo}, \& {Park}}]{choi2009}
{Choi}, Y.-Y., {Woo}, J.-H., \& {Park}, C. 2009, \apj, 699, 1679


\bibitem[{{Cisternas} {et~al.}(2011){Cisternas}, {Jahnke}, {Inskip},
  {Kartaltepe}, {Koekemoer}, {Lisker}, {Robaina}, {Scodeggio}, {Sheth},
  {Trump}, {Andrae}, {Miyaji}, {Lusso}, {Brusa}, {Capak}, {Cappelluti},
  {Civano}, {Ilbert}, {Impey}, {Leauthaud}, {Lilly}, {Salvato}, {Scoville}, \&
  {Taniguchi}}]{cisternas2011}
{Cisternas}, M., {Jahnke}, K., {Inskip}, K.~J., {et~al.} 2011, \apj, 726, 57

\bibitem[{{Comastri} {et~al.}(2011){Comastri}, {Ranalli}, {Iwasawa}, {Vignali},
  {Gilli}, {Georgantopoulos}, {Barcons}, {Brandt}, {Brunner}, {Brusa},
  {Cappelluti}, {Carrera}, {Civano}, {Fiore}, {Hasinger}, {Mainieri},
  {Merloni}, {Nicastro}, {Paolillo}, {Puccetti}, {Rosati}, {Silverman},
  {Tozzi}, {Zamorani}, {Balestra}, {Bauer}, {Luo}, \& {Xue}}]{comastri2011}
{Comastri}, A., {Ranalli}, P., {Iwasawa}, K., {et~al.} 2011, \aap, 526, L9

\bibitem[{{Connelly} {et~al.}(2012){Connelly}, {Wilman}, {Finoguenov}, {Hou},
  {Mulchaey}, {McGee}, {Balogh}, {Parker}, {Saglia}, {Henderson}, \&
  {Bower}}]{connelly2012}
{Connelly}, J.~L., {Wilman}, D.~J., {Finoguenov}, A., {et~al.} 2012, \apj, 756,
  139

\bibitem[{{Cooper} {et~al.}(2012){Cooper}, {Yan}, {Dickinson}, {Juneau},
  {Lotz}, {Newman}, {Papovich}, {Salim}, {Walth}, {Weiner}, \&
  {Willmer}}]{cooper2012}
{Cooper}, M.~C., {Yan}, R., {Dickinson}, M., {et~al.} 2012, \mnras, 425, 2116

\bibitem[Dahlen et al.(2005)]{dahlen2005} Dahlen, T., Mobasher, 
B., Somerville, R.~S., et al.\ 2005, \apj, 631, 126 

\bibitem[Digby-North et al.(2010)]{digby2010} Digby-North, J.~A., 
Nandra, K., Laird, E.~S., et al.\ 2010, \mnras, 407, 846 

\bibitem[{{Eastman} {et~al.}(2007){Eastman}, {Martini}, {Sivakoff}, {Kelson},
  {Mulchaey}, \& {Tran}}]{eastman2007}
{Eastman}, J., {Martini}, P., {Sivakoff}, G., {et~al.} 2007, \apjl, 664, L9


\bibitem[Elbaz et al.(2011)]{elbaz2011} Elbaz, D., Dickinson, M., Hwang, H.~S., et al.\ 2011, \aap, 533, A119 

\bibitem[Finoguenov et al.(2009)]{fin09} Finoguenov, A., 
Connelly, J.~L., Parker, L.~C., et al.\ 2009, \apj, 704, 564

\bibitem[Finoguenov et al.(2010)]{fin10} Finoguenov, A., 
Watson, M.~G., Tanaka, M., et al.\ 2010, \mnras, 403, 2063


\bibitem[Finoguenov et al. (2014)]{finoguenov2014}
{Finoguenov}, A., Tanaka, M., Cooper, M., Allevato, V., Cappelluti, N., Choi, A, Heymans, C., Bauer, F.E., et al. 2014, 
\aap, submitted 

\bibitem[{{Galametz} {et~al.}(2009){Galametz}, {Stern}, {Eisenhardt},
  {Brodwin}, {Brown}, {Dey}, {Gonzalez}, {Jannuzi}, {Moustakas}, \&
  {Stanford}}]{galametz2009}
{Galametz}, A., {Stern}, D., {Eisenhardt}, P.~R.~M., {et~al.} 2009, \apj, 694,
  1309

\bibitem[Gabor et al.(2009)]{garbor2009} Gabor, J.~M., Impey, 
C.~D., Jahnke, K., et al.\ 2009, \apj, 691, 705 

\bibitem[Georgakakis et al.(2008)]{georga2008} Georgakakis, A., 
Gerke, B.~F., Nandra, K., et al.\ 2008, \mnras, 391, 183


\bibitem[{{Gehren} {et~al.}(1984){Gehren}, {Fried}, {Wehinger}, \&
  {Wyckoff}}]{gehren1984}
{Gehren}, T., {Fried}, J., {Wehinger}, P.~A., \& {Wyckoff}, S. 1984, \apj, 278,
  11

\bibitem[{{Grogin} {et~al.}(2005){Grogin}, {Conselice}, {Chatzichristou},
  {Alexander}, {Bauer}, {Hornschemeier}, {Jogee}, {Koekemoer}, {Laidler},
  {Livio}, {Lucas}, {Paolillo}, {Ravindranath}, {Schreier}, {Simmons}, \&
  {Urry}}]{grogin2005}
{Grogin}, N.~A., {Conselice}, C.~J., {Chatzichristou}, E., {et~al.} 2005,
  \apjl, 627, L97

\bibitem[Haggard et al.(2010)]{haggard2010} Haggard, D., Green, 
P.~J., Anderson, S.~F., et al.\ 2010, \apj, 723, 1447 

\bibitem[H{\"a}ussler et al.(2007)]{haussler2007} H{\"a}ussler, B., 
McIntosh, D.~H., Barden, M., et al.\ 2007, \apjs, 172, 615 

\bibitem[{{Heckman} {et~al.}(1984){Heckman}, {Bothun}, {Balick}, \&
  {Smith}}]{heckman1984}
{Heckman}, T.~M., {Bothun}, G.~D., {Balick}, B., \& {Smith}, E.~P. 1984, \aj,
  89, 958

\bibitem[{{Hennawi} {et~al.}(2006){Hennawi}, {Strauss}, {Oguri}, {Inada},
  {Richards}, {Pindor}, {Schneider}, {Becker}, {Gregg}, {Hall}, {Johnston},
  {Fan}, {Burles}, {Schlegel}, {Gunn}, {Lupton}, {Bahcall}, {Brunner}, \&
  {Brinkmann}}]{hennawi2006}
{Hennawi}, J.~F., {Strauss}, M.~A., {Oguri}, M., {et~al.} 2006, \aj, 131, 1

\bibitem[Hopkins et al.(2010)]{hopkins2010a} Hopkins, P.~F., Bundy, 
K., Croton, D., et al.\ 2010, \apj, 715, 202 

\bibitem[{{Hutchings} \& {Neff}(1992)}]{hutchings1992}
{Hutchings}, J.~B., \& {Neff}, S.~G. 1992, \aj, 104, 1

\bibitem[Ilbert et 
al.(2005)]{ilbert2005} Ilbert, O., Tresse, L., Zucca, E., et al.\ 2005, \aap, 439, 863 

\bibitem[Kocevski et al.(2012)]{kocevski2012} Kocevski, D.~D., 
Faber, S.~M., Mozena, M., et al.\ 2012, \apj, 744, 148 


\bibitem[Koss et al.(2011)]{koss2011} Koss, M., Mushotzky, R., Veilleux, S., et al.\ 2011, \apj, 739, 57 

\bibitem[Lee et al.(2012)]{lee2012} Lee, G.-H., Woo, J.-H., 
Lee, M.~G., et al.\ 2012, \apj, 750, 141 

\bibitem[Lehmer et al.(2009)]{lehmer2009} Lehmer, B.~D., 
Alexander, D.~M., Geach, J.~E., et al.\ 2009, \apj, 691, 687 

\bibitem[Lutz et al.(2011)]{lutz2011} Lutz, D., Poglitsch, A., Altieri, B., et al.\ 2011, \aap, 532, A90 

\bibitem[{{Martini}(2004)}]{martini2004}
{Martini}, P. 2004, in IAU Symposium, Vol. 222, The Interplay Among Black
  Holes, Stars and ISM in Galactic Nuclei, ed. T.~{Storchi-Bergmann}, L.~C.
  {Ho}, \& H.~R. {Schmitt}, 235--241

\bibitem[{{Martini} {et~al.}(2007){Martini}, {Mulchaey}, \&
  {Kelson}}]{martini2007}
{Martini}, P., {Mulchaey}, J.~S., \& {Kelson}, D.~D. 2007, \apj, 664, 761

\bibitem[{{Martini} {et~al.}(2009){Martini}, {Sivakoff}, \&
  {Mulchaey}}]{martini2009}
{Martini}, P., {Sivakoff}, G.~R., \& {Mulchaey}, J.~S. 2009, \apj, 701, 66

\bibitem[Martini et al.(2013)]{martini2013} Martini, P., Miller, 
E.~D., Brodwin, M., et al.\ 2013, \apj, 768, 1 

\bibitem[{{Mulchaey} {et~al.}(2006){Mulchaey}, {Lubin}, {Fassnacht}, {Rosati},
  \& {Jeltema}}]{mulchaey2006}
{Mulchaey}, J.~S., {Lubin}, L.~M., {Fassnacht}, C., {Rosati}, P., \& {Jeltema},
  T.~E. 2006, \apj, 646, 133

\bibitem[Mullaney et al.(2011)]{mullaney2011} Mullaney, J.~R., 
Alexander, D.~M., Goulding, A.~D., \& Hickox, R.~C.\ 2011, \mnras, 414, 1082 

\bibitem[Mullaney et al.(2012)]{mullaney2012a} Mullaney, J.~R., 
Pannella, M., Daddi, E., et al.\ 2012, \mnras, 419, 95 

\bibitem[Pentericci et al.(2013)]{pentericci2013} Pentericci, L., Castellano, M., Menci, N., et al.\ 2013, \aap, 552, A111

\bibitem[Rigby et al.(2009)]{rigby2009} Rigby, J.~R., Diamond-Stanic, A.~M., \& Aniano, G.\ 2009, \apj, 700, 1878 

\bibitem[{{Rix} {et~al.}(2004){Rix}, {Barden}, {Beckwith}, {Bell}, {Borch},
  {Caldwell}, {H{\"a}ussler}, {Jahnke}, {Jogee}, {McIntosh}, {Meisenheimer},
  {Peng}, {Sanchez}, {Somerville}, {Wisotzki}, \& {Wolf}}]{rix2004}
{Rix}, H.-W., {Barden}, M., {Beckwith}, S.~V.~W., {et~al.} 2004, \apjs, 152,
  163

\bibitem[Rosario et al.(2012)]{rosario2012} Rosario, D.~J., Santini, P., Lutz, D., et al.\ 2012, \aap, 545, A45 

\bibitem[{{Ruderman} \& {Ebeling}(2005)}]{ruderman2005}
{Ruderman}, J.~T., \& {Ebeling}, H. 2005, \apjl, 623, L81

\bibitem[{{Serber} {et~al.}(2006){Serber}, {Bahcall}, {M{\'e}nard}, \&
  {Richards}}]{serber2006}
{Serber}, W., {Bahcall}, N., {M{\'e}nard}, B., \& {Richards}, G. 2006, \apj,
  643, 68

\bibitem[{{Shen} {et~al.}(2007){Shen}, {Mulchaey}, {Raychaudhury}, {Rasmussen},
  \& {Ponman}}]{shen2007}
{Shen}, Y., {Mulchaey}, J.~S., {Raychaudhury}, S., {Rasmussen}, J., \&
  {Ponman}, T.~J. 2007, \apjl, 654, L115


\bibitem[{{Silverman} {et~al.}(2009){Silverman}, {Kova{\v c}}, {Knobel},
  {Lilly}, {Bolzonella}, {Lamareille}, {Mainieri}, {Brusa}, {Cappelluti},
  {Peng}, {Hasinger}, {Zamorani}, {Scodeggio}, {Contini}, {Carollo}, {Jahnke},
  {Kneib}, {Le Fevre}, {Bardelli}, {Bongiorno}, {Brunner}, {Caputi}, {Civano},
  {Comastri}, {Coppa}, {Cucciati}, {de la Torre}, {de Ravel}, {Elvis},
  {Finoguenov}, {Fiore}, {Franzetti}, {Garilli}, {Gilli}, {Griffiths},
  {Iovino}, {Kampczyk}, {Koekemoer}, {Le Borgne}, {Le Brun}, {Maier},
  {Mignoli}, {Pello}, {Perez Montero}, {Ricciardelli}, {Tanaka}, {Tasca},
  {Tresse}, {Vergani}, {Vignali}, {Zucca}, {Bottini}, {Cappi}, {Cassata},
  {Marinoni}, {McCracken}, {Memeo}, {Meneux}, {Oesch}, {Porciani}, \&
  {Salvato}}]{silverman2009}
{Silverman}, J.~D., {Kova{\v c}}, K., {Knobel}, C., {et~al.} 2009, \apj, 695,
  171

\bibitem[{{Silverman} {et~al.}(2010){Silverman}, {Mainieri}, {Salvato},
  {Hasinger}, {Bergeron}, {Capak}, {Szokoly}, {Finoguenov}, {Gilli}, {Rosati},
  {Tozzi}, {Vignali}, {Alexander}, {Brandt}, {Lehmer}, {Luo}, {Rafferty},
  {Xue}, {Balestra}, {Bauer}, {Brusa}, {Comastri}, {Kartaltepe}, {Koekemoer},
  {Miyaji}, {Schneider}, {Treister}, {Wisotski}, \& {Schramm}}]{silverman2010}
{Silverman}, J.~D., {Mainieri}, V., {Salvato}, M., {et~al.} 2010, \apjs, 191,
  124

\bibitem[Silverman et al.(2011)]{2011ApJ...743....2S} Silverman, J.~D., 
Kampczyk, P., Jahnke, K., et al.\ 2011, \apj, 743, 2 

\bibitem[Tanaka et al.(2012)]{tanaka12} Tanaka, M., Finoguenov, 
A., Lilly, S.~J., et al.\ 2012, \pasj, 64, 22

\bibitem[Tanaka et al.(2013)]{tanaka2013} Tanaka, M., Finoguenov, 
A., Mirkazemi, M., et al.\ 2013, \pasj, 65, 17

\bibitem[Ueda et al.(2003)]{ueda2003} Ueda, Y., Akiyama, M., 
Ohta, K., \& Miyaji, T.\ 2003, \apj, 598, 886 

\bibitem[{{Xue} {et~al.}(2011){Xue}, {Luo}, {Brandt}, {Bauer}, {Lehmer},
  {Broos}, {Schneider}, {Alexander}, {Brusa}, {Comastri}, {Fabian}, {Gilli},
  {Hasinger}, {Hornschemeier}, {Koekemoer}, {Liu}, {Mainieri}, {Paolillo},
  {Rafferty}, {Rosati}, {Shemmer}, {Silverman}, {Smail}, {Tozzi}, \&
  {Vignali}}]{xue2011}
{Xue}, Y.~Q., {Luo}, B., {Brandt}, W.~N., {et~al.} 2011, \apjs, 195, 10

\bibitem[{{Yan} {et~al.}(2011){Yan}, {Ho}, {Newman}, {Coil}, {Willmer},
  {Laird}, {Georgakakis}, {Aird}, {Barmby}, {Bundy}, {Cooper}, {Davis},
  {Faber}, {Fang}, {Griffith}, {Koekemoer}, {Koo}, {Nandra}, {Park},
  {Sarajedini}, {Weiner}, \& {Willner}}]{yan2011}
{Yan}, R., {Ho}, L.~C., {Newman}, J.~A., {et~al.} 2011, \apj, 728, 38

\bibitem[Ziparo et al.(2013)]{ziparo2013} Ziparo, F., Popesso, P., 
Biviano, A., et al.\ 2013, \mnras, 1855 

\end{thebibliography}
\end{document}